\documentclass[reprint, amsmath,amssymb,aps]{revtex4-2}
\usepackage{bbm}
\usepackage{graphicx}

    \newcommand{\ket}[1]{\left| #1 \right\rangle}
    \newcommand{\bra}[1]{\left\langle #1 \right|}

    \newcommand{\Tr}{\text{Tr}}

\begin{document}

\title{Qubit Quantum Metrology with Limited Measurement Resources}
\author{Jason Saunders}
	\email{jason.saunders@byu.net}
\author{Jean-Francois Van Huele}
	\email{vanhuele@byu.edu}
	\affiliation{Department of Physics \& Astronomy, Brigham Young University, Provo, Utah 84602}
\date{\today}

\begin{abstract}
	Quantum resources, such as entanglement, can decrease the uncertainty of a parameter-estimation procedure beyond what is classically possible. This phenomenon is well described for noiseless systems with asymptotically many measurement resources by the Quantum Cramer-Rao Bound, but no general description exists for the regime of limited measurement resources. We address this problem by defining a Bayesian quantifier for uncertainty suitable for the regime of limited resources, and by developing a mathematical description for a parameter-estimation procedure which uses qubit probes to estimate a rotation angle induced on them. We simulate the qubit estimation scheme in the regime of limited resources using a single class of probe states. We find that, in noiseless systems, entanglement between qubits always decreases the uncertainty of the estimation; however, the quantum advantage decreases as fewer qubits are used in the estimation. We also find that the presence of strong dephasing noise removes the quantum advantage completely, regardless of the number of qubit probes used.
\end{abstract}

\maketitle



\section{The Quantum Advantage: How Quantum Resources Can Decrease Uncertainty} \label{sec:introduction}

Parameter estimation plays a central role in science. Anytime we ask a quantitative question, like ``What is the temperature of the substance?'', ``When did the event happen?'', or ``Where is the system located?'', we need to estimate the relevant parameter in order to answer our question. The more precise our estimation is, the more specific of a question we can answer. Improving our estimations has the potential to benefit scientific investigation across many fields. Metrology is precisely the study of parameter estimation and how to improve it, and quantum metrology investigates how quantum resources, such as entanglement, can be used to improve estimations. In this work, we will investigate how utilizing entanglement in particular can decrease the uncertainty of a parameter estimation.

The terms parameter estimation and measurement are often used interchangeably, so it is worth distinguishing between the two for the purposes of this work. A \emph{parameter estimation} is the entire process of determining the value of some parameter of interest. This involves preparing probes, interacting them with the system of interest, measuring the probes, then taking the measurement results and constructing an estimate for the parameter of interest. A \emph{measurement}, then, is a step in the estimation procedure. It involves extracting the relevant information from the probes after they have interacted with the system. A parameter estimation may include any number of measurements. 


Quantum resources can efficiently perform classically challenging--or in some cases classically impossible--tasks. Some examples of these tasks include using quantum computers to solve specific problems orders of magnitude faster than classical computers, or using quantum cryptography to achieve unbreakable connections.

\emph{Quantum resources} and \emph{quantum advantage} are central to this work and the field of quantum information at large. Quantum mechanics describes phenomena with no classical parallel--some examples include superposition, particle tunneling, and entanglement. These phenomena are the keys which allow us to perform the aforementioned classically-challenging or classically-impossible tasks. When a quantum phenomenon is utilized in a task, we label it a quantum resource. The quantum procedure described in this work uses quantum entanglement as its quantum resource. Whenever a procedure utilizing quantum resources is able to outperform a similar classical procedure, we say that we have achieved a quantum advantage--the use of quantum resources has given us an advantage. As demonstrated by the Quantum Cramer-Rao Bound \cite{BraunsteinCaves1994}, we can achieve a quantum advantage in the field of metrology.

For noiseless systems, the Quantum Cramer-Rao Bound describes the lowest achievable uncertainty of an estimation procedure. It is a quantum extension of a classical result, the Cramer-Rao Bound \cite{Cramer1946}. The Quantum Cramer-Rao Bound states
\begin{align}
	\delta \phi \geq \frac{1}{(\Delta h)(2 \sqrt{\nu})}.
\end{align}
In this inequality $\delta \phi$ is the uncertainty of $\phi$, the parameter we are estimating. The quantity $\Delta h$ is the variance of $h$, the generator of $\phi$. The generator $h$ encodes the interaction which imprints $\phi$ onto our measurement probes. Finally, $\nu$ is the number of measurements we make in our estimation. This value is equivalent to the number of probes used in the estimation, assuming each probe is measured once. The Quantum Cramer-Rao Bound assumes that the number of measurements $\nu$ is asymptotically large. In this regime, the Quantum Cramer-Rao Bound is tight, meaning that there exists some system which can reach the lower bound.

The Quantum Cramer-Rao Bound is exact when $\nu$ is infinitely large. As we explore the more realistic regime of finite $\nu$, the error between the Quantum Cramer-Rao Bound and the true lower bound should become infinitesimally small as the number of measurements approaches infinity. We consider the number of measurements to be \emph{asymptotically large} when the difference between the true lower bound and the lower bound predicted by the Quantum Cramer-Rao Bound is negligible. What is considered negligible depends on the level of precision required, so there is not a unique boundary between the asymptotic regime and its counter part, the limited or non-asymptotic regime. For this work, we define the \emph{limited regime} as the area where results differ qualitatively from the asymptotic regime.

The Quantum Cramer-Rao Bound is a very general condition; it can be derived for specific systems by choosing the right generator $h$. In \cite{Giovannetti2006}, this bound is derived for a more specific estimation procedure: $N$ identical probes per measurement. The lower bound of the uncertainty $\delta \phi_s$ is derived assuming that the $N$ probes are separable (uncorrelated), and the lower bound $\delta \phi_e$ is derived by assuming that all $N$ probes are maximally entangled. Comparing the two bounds, the authors find
\begin{align} \label{eq:relative_qcrb}
	\delta \phi_e = \frac{\delta \phi_s}{\sqrt{N}}.
\end{align}
For this class of systems, entanglement reduces the lower bound of the uncertainty by a factor of $1/\sqrt{N}$ in the regime of asymptotically large $\nu$.

The Quantum Cramer-Rao Bound only holds when the number of measurements $\nu$ is asymptotically large. This covers a large space of possible estimation scenarios, but there exist other situations where the number of measurements is small and limited. Imagine that an estimation procedure is destructive, so a sample can only be measured a handful of times. Or perhaps the measurement target is in motion, and there is a limited time window available for measurement. In these cases, we want to understand how quantum resources can improve our estimation procedures. Is the benefit of entanglement the same in the limited regime as in the asymptotic regime? Is the quantum advantage larger, smaller, or perhaps nonexistent?

This non-asymptotic, or limited, regime has been studied before. Some analytical quantum bounds for the uncertainty have been proposed, such as the Ziv-Zakai bound \cite{Tsang2012} and the Weiss-Weinstein bound \cite{LuTsang2016}, but neither of these bounds are tight--we are not guaranteed to be able to reach their lower bounds \cite{Rubio2019}. Specific systems, such as interferometric estimation schemes, have also been investigated numerically in this regime \cite{Rubio2019}, but the results do not extend to other estimation procedures.

The discussion thus far has assumed the absence of noise, but noise is an important factor to consider when designing real-world applications. In the asymptotic regime, it has been shown that the presence of even small amounts of some types of noise, such as dephasing, can completely wipe out the quantum advantage, although there is still a reduced quantum advantage in the presence of other types of noise \cite{DemkowiczDobrzanski2014}. It has also been shown that the quantum advantage can be salvaged if the number of measurements is kept small enough \cite{Nichols2016}.



In this work, we study a specific estimation procedure in the non-asymptotic regime: estimating a rotation induced on qubit probes. In Sec. \ref{sec:bayesian_quantifier}, we use Bayesian methods to develop a measure of uncertainty which is appropriate for regimes of very limited measurement information. We develop the equations and transformations necessary to simulate the qubit estimation protocol in Sec. \ref{sec:qubit_formalism}. We additionally include the transformations necessary to simulate dephasing noise in the qubits. Important computational considerations are detailed in Sec. \ref{sec:comp_considerations}, and results for the qubit simulations are presented in Sec. \ref{sec:results}. We use a single class of states as probes to study the effect that entanglement between the qubits has on the uncertainty of the estimation. We  also simulate dephasing noise, a type of decoherence, and determine how it affects the quantum advantage. We find that, in the noiseless simulations, entanglement provides a quantum advantage, although the advantage decreases when fewer probes are used. The presence of weak dephasing noise reduces this quantum advantage, but entanglement still decreases the uncertainty of the estimation. Strong dephasing noise removes the quantum advantage entirely. We conclude in Sec. \ref{sec:conclusion}.

In the current work we deal exclusively with a qubit metrology scheme. Similar numerical approaches have been used to study interferometric metrology schemes in the regime of limited resources \cite{Rubio2019, MySeniorThesis}.



\section{Measurement and Parameter Estimation in the Regime of Limited Resources} \label{sec:bayesian_quantifier}

\subsection{Motivating a Bayesian Approach}

Estimating a parameter with limited resources is challenging. Imagine first the following situation: we want to measure the angle of a magnetic field using many neutrally charged spin particles, initialized in the spin up position. We  send the particles through the field, then measure each one individually, recording either spin up or spin down. From our measurement results we can reconstruct the wave function of the particles, from which we can estimate the angle of the magnetic field. We  could further define an uncertainty of our estimation by repeating the procedure many times and recording the variation in the estimation result.

Now imagine this second situation: we again want to measure the angle of a magnetic field, but this time we have only a single neutrally charged spin particle. We  initialize the particle in the spin up position and send it through the field. We  measure it, recording either spin up or spin down. From this single datapoint, what can we say about the angle of the magnetic field? We lack the information necessary to reconstruct the wave function with any degree of certainty, and so would be hard pressed to come up with an angle of the field, or an uncertainty of our estimation, using the same methods as we used in the many particle case. So what can we say?

\subsection{A Bayesian Quantification of the Uncertainty} \label{sec:bayesian_uncertainty}

Bayesian methods center on the idea of calculating the probability, or likelihood, of a hypothesis. Given our measurement results, whether they be a single data point or many, we can calculate the probability that the parameter of interest lies within a specific range. We  can use Bayesian ideas to define an uncertainty metric for our limited-resources estimation procedure. To do this, we will first introduce some useful functions, then use them to quantify the uncertainty of a general parameter $\phi$ in the regime of a non-asymptotic, limited number of measurements $\nu$. 

\begin{figure}
    \centerline{\includegraphics[width=.75\linewidth]{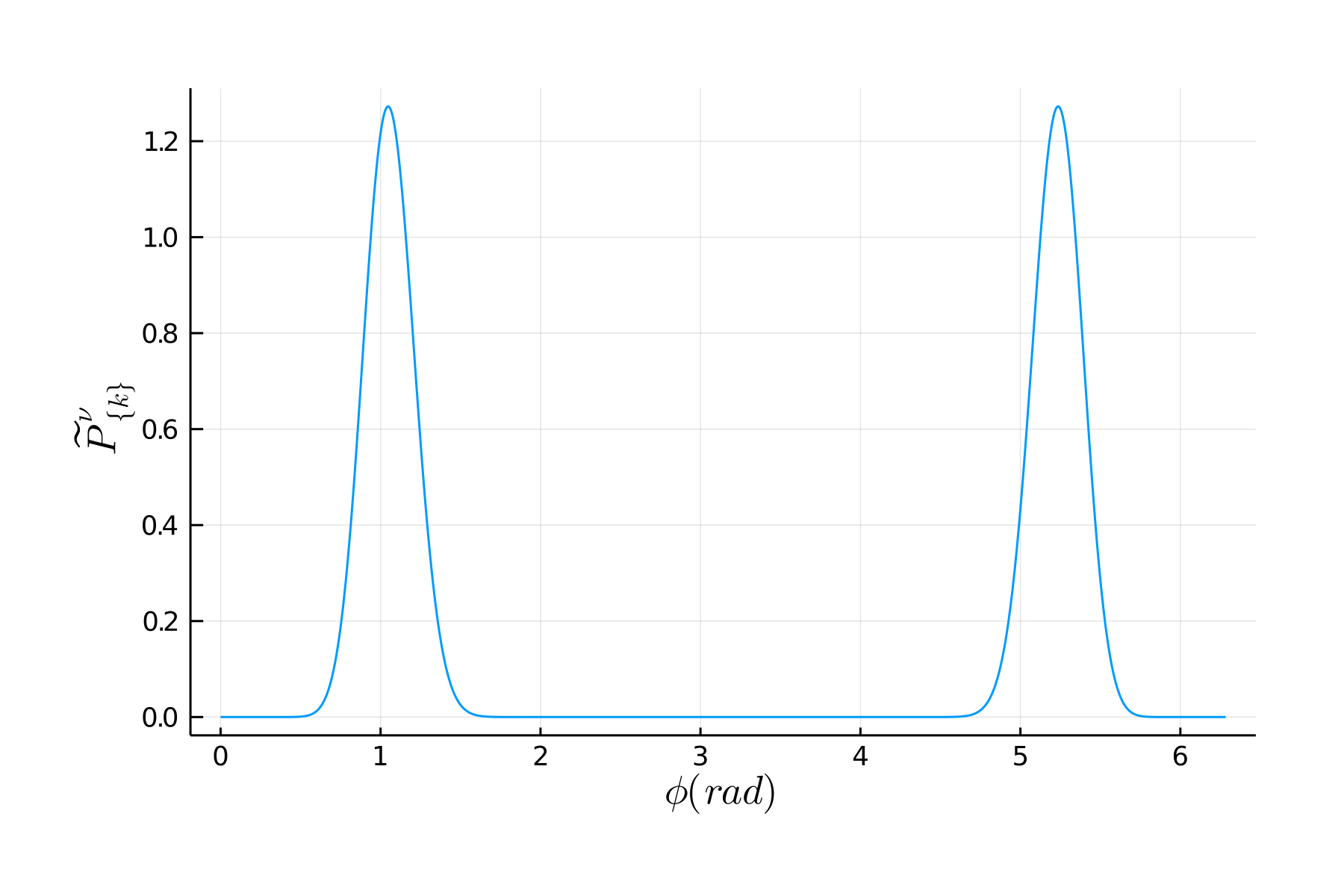}}
    \caption{\label{fig:phi_mp_10} Example of a possible probability distribution $\widetilde{P}_{\{k\}}^\nu (\phi)$ for a set of $\nu$ measurements $\{k\}$ generated using the methods described in Sec. \ref{sec:qubit_formalism}.}
\end{figure}

First we will define a complete, orthogonal measurement basis, comprised of states $\ket{\xi_1}, \dots, \ket{\xi_m}$, where $m$ is the dimensionality of the probe. Let $\Xi_i (\phi)$ be the probability of measuring $\ket{\xi_i}$ if the initial state $\ket{\psi}$ has been transformed by the $\phi$-dependent evolution $U_\phi$	
	\begin{equation} \label{eq:Xi}
		\Xi_i \left(\phi \right)= \left| \bra{\xi_i} U_\phi\left(\phi\right) \ket{\psi} \right|^2.
	\end{equation}
Next, we'll define a function $P_{\{k\}}^\nu (\phi, \nu, k_1, k_2, \dots, k_m)$, which is an unnormalized probability distribution as a function of the parameter $\phi$, given that $\ket{\xi_1}$ was measured $k_1$ times, $\ket{\xi_2}$ $k_2$ times, and so on, for a total of $\nu$ measurements:
	\begin{equation}
		P_{\{k\}}^\nu \left(\phi, \nu, k_1, \dots, k_m \right) = \frac{\nu!}{k_1!k_2!\cdots k_m!} \prod_{i=1}^m \Xi_i^{k_i} (\phi).
	\end{equation}
	The fraction on the right-hand side of the equation is a combinatorial factor resulting from counting all the possible orderings of the $\nu$ measurements, which is multiplied by the probability of measuring that set of measurements. Finally, we define a third function, a weighted, normalized probability distribution,
	\begin{equation} \label{eq:Pnk_tilde}
		\widetilde{P}_{\{k\}}^\nu = \frac{1}{A} w(\phi) P_{\{k\}}^\nu(\phi, \nu, \{k\}).
	\end{equation}
The function $w(\phi)$ is a weight factor, allowing us to encode prior-known information about the probabilities of different values of the parameter $\phi$. For example, if all $\phi$ values are equally likely, then $w(\phi) = 1$. The variable $A$ is a normalization constant, and is defined by the equation
	\begin{equation}
		A = \int w(\phi) P_{\{k\}}^\nu(\phi, \nu, \{k\}) \, d\phi.
	\end{equation}
Plotting this function, $\widetilde{P}_{\{k\}}^\nu (\phi, \nu, \{k\})$, against the parameter $\phi$ will give us some functional dependence, as seen in the example from Fig. \ref{fig:phi_mp_10}.

	We can draw a few conclusions from this plot. First, notice that there is an intrinsic width or range to the probe. In this example, it cannot distinguish between parameters $\phi = \pi - x$ and $\phi = \pi + x$, where $x \in [0, \pi]$. This range will differ for different systems and different probes, but it suffices to say that we need to restrict our analysis to the range of sensitivity of the probe. In this case, we will restrict further analysis to the range $\phi \in [0, \pi]$. In this restricted range, we define $\phi_{mp}$, the most probable value of $\phi$ given the information we have. The quantity $\phi_{mp}$ is defined by the equation
	\begin{equation} \label{eq:phi_mp}
		\widetilde{P}_{\{k\}}^\nu(\phi_{mp}, \nu, \{k\}) = \max_\phi \left[\widetilde{P}_{\{k\}}^\nu(\phi, \nu, \{k\}) \right].
	\end{equation}
	We can also define the probability that the parameter $\phi$ is between $a$ and $b$, with $a$ and $b$ in the domain of $\phi$:
	\begin{equation}
		P(a,b) = \int_a^b \widetilde{P}_{\{k\}}^\nu(\phi, \nu, \{k\}) \, d\phi.
	\end{equation}
	Another useful value is the minimum length of the confidence interval $y$, defined as
	\begin{equation}
		L(CI)_y = \min_{a, b \in D(\phi)}(b - a),
	\end{equation}
	such that
	\begin{equation}
		y = \int_a^b \widetilde{P}_{\{k\}}^\nu(\phi, \nu, \{k\}) \, d\phi.
	\end{equation}
	The symbol $D(\phi)$ denotes the domain of $\phi$. Notice that while multiple confidence intervals are possible, this expression gives us the smallest length necessary to achieve a confidence of $y$.
	
	There is one more important element to this framework. Each estimation has the potential to be different, due to the random nature of measurement. To compensate for this fact, we will simulate a large number of estimations, quantified in the variable $N_E$, with the same $\nu$ and initial probe state, then compute four metrics from this set, which we define as $\mu (\phi_{mp}), \sigma (\phi_{mp}), \mu (L(CI)_{.95}),$ and  $\sigma (L(CI)_{.95})$. If $\phi_{mp, i}$ and $L(CI)_{.95, i}$ are the $\phi_{mp}$  and $L(CI)_{.95}$ estimated by the $i$th estimation in the set of $N_E$ estimations, then our metrics are defined as follows:
	
	\begin{align}
		\mu (\phi_{mp}) &= \frac{1}{N_E} \sum_i^{N_E} \phi_{mp, i} \\
    		\sigma (\phi_{mp}) &= \sqrt{\frac{\sum_i^{N_E} \left(\phi_{mp, i} - \mu(\phi_{mp}) \right)^2}{N_E-1}} \\
   		\mu (L(CI)_{.95}) &= \frac{1}{N_E} \sum_i^{N_E} L(CI)_{.95, i} \\
    		\sigma (L(CI)_{.95}) &= \sqrt{\frac{\sum_i^{N_E} \left(L(CI)_{.95, i}- \mu(L(CI)_{.95}) \right)^2}{N_E-1}}.
	\end{align}
	
	The quantities $\mu (\phi_{mp})$ and $\sigma (\phi_{mp})$ are the mean and standard deviation of the $\phi_{mp}$ found by each estimation in the set, and $\mu (L(CI)_{.95})$ and  $\sigma (L(CI)_{.95})$ are the mean and standard deviation of the length of the 95\% confidence interval found by each estimation in the set. The 95\% confidence value was chosen somewhat arbitrarily as a value with good certainty, but not so high that it would always span the majority of the domain of $\phi$. In a nutshell, the parameters $\mu (\phi_{mp})$ and $\mu (L(CI)_{.95})$ encode the average accuracy and uncertainty of an estimation, and $\sigma (\phi_{mp})$ and $\sigma (L(CI)_{.95})$ encode how much variation there is in the accuracy and uncertainty between estimations. An ideal measurement set would find $\mu (\phi_{mp}) = \phi$, $\sigma ( \phi_{mp} ) = 0$, $\mu (L(CI)_{.95}) = 0$ and  $\sigma (L(CI)_{.95}) = 0$. Our analysis will use the metric $\mu(L(CI))$ as our uncertainty metric. 
	
	The metric $\mu(L(CI))$ (as well as the other three metrics) are all dependent on the parameter $\phi$. We assume that we have no prior knowledge about the parameter $\phi$, so we are interested in reducing the uncertainty averaged over all possible angles $\phi$. We will denote this average using the notation mean($\mu(L(CI))$).
	
	Now that we have defined an uncertainty metric appropriate for the regime of limited resources, we will formally introduce our qubit estimation procedure, and calculate the uncertainty for the setup.


\section{Qubit Metrology Scheme} \label{sec:qubit_formalism}

\subsection{General Qubit Metrology Scheme}

A qubit is any quantum state in a 2-dimensional space. Examples include electron-spin states and two-level atoms. Unlike a classical bit, which only has two possible values, a qubit can exist in infinitely many different superpositions of its two basis states. In our estimation scheme, we will initialize qubit probes into a known state, then induce a rotation on the probes by an unknown angle $\phi$, and then estimate the angle of rotation using the methods described in the previous section.

In order to allow for entanglement within the probes, each probe will be comprised of two qubits. A single probe state, $\ket{\psi}$, will be represented as
\begin{align}
	\ket{\psi} = \alpha \ket{\downarrow_1 \downarrow_2} + \beta \ket{\downarrow_1 \uparrow_2} + \gamma \ket{\uparrow_1 \downarrow_2} + \delta \ket{\uparrow_1 \uparrow_2} = \begin{pmatrix} \alpha \\ \beta \\ \gamma \\ \delta \end{pmatrix},
\end{align}
where we have used the tensor-product representation of the two qubits. The state is normalized, so $|\alpha|^2 + |\beta|^2 + |\gamma|^2 + |\delta|^2 = 1$, and the subscripts 1 and 2 refer to the first and second qubit respectively, although they will be omitted in what follows. We  are specifically interested in the effect of entanglement on the uncertainty of our estimation, so we will test initial probe states of the form
\begin{align}
	\ket{\psi} = \sqrt{\alpha} \ket{\downarrow \uparrow} + \sqrt{1-\alpha} \ket{\uparrow \downarrow} =  \begin{pmatrix} 0 \\ \sqrt{\alpha} \\ \sqrt{1-\alpha} \\ 0 \end{pmatrix}.
\end{align}
The parameter $\alpha$ is chosen to be real and varies in the range $[0, 1/2]$, with $\alpha=0$ corresponding to a separable, non-entangled state, and $\alpha=1/2$ corresponding to the Bell state $\ket{\Psi^+}$, a maximally entangled state. We assume no prior information about the parameter $\phi$, so in the notation of Sec. \ref{sec:bayesian_uncertainty} we have $w(\phi) = 1$. The rotation $U_\phi$ will act independently on each qubit within the probe, and is expressed by the tensor-product matrix
\begin{align} 
	U_\phi = \begin{pmatrix} \cos (\phi/2) & \sin (\phi/2) \\ -\sin (\phi/2) & \cos(\phi/2) \end{pmatrix} \otimes \begin{pmatrix} \cos (\phi/2) & \sin (\phi/2) \\ -\sin (\phi/2) & \cos(\phi/2) \end{pmatrix}.
\end{align}
An initial probe state $\ket{\psi_i}$ is evolved into the final probe state $\ket{\psi_f}$ by the formula
\begin{align}
	\ket{\psi_f} = U_\phi \ket{\psi_i}.
\end{align}

These states and evolutions which we have introduced have all been for pure quantum states not affected by noise. However, we are also interested in the effect of dephasing on our qubit estimation procedure, and so will now generalize the previous formulae to include mixed, or noisy, quantum states as well. As the name suggests, a mixed quantum state $\rho$ can be thought of as a mixture of different pure quantum states, represented mathematically as
\begin{align}
	\rho = \sum_i p_i \ket{\psi_i}\bra{\psi_i}.
\end{align}
The $\ket{\psi_i}$ are pure states, and the weight factor $p_i$ can be thought of as the probability that a mixed state $\rho$ will be measured in the state $\ket{\psi_i}$, although the expansion of a mixed state into a sum of pure states is not unique. To rotate an initial mixed state $\rho_i$ by an angle $\phi$, the formula becomes
\begin{align}
	\rho_f = U_\phi \rho_i U_\phi^\dagger,
\end{align}
where the rotation matrix $U_\phi$ is unchanged from the pure state case.

The function $\Xi_i(\phi)$, as introduced in Sec. \ref{sec:bayesian_uncertainty} and from which we can calculate the $L(CI)$ for an estimation, is
\begin{align}
	\Xi_i (\phi) = | \bra{\xi_i} U_\phi \ket{\psi}|^2 = \Tr\left(\ket{\xi_i}\bra{\xi_i} U_\phi \rho U_\phi^\dagger \right).
\end{align}
The possible measurement states $\ket{\xi_i}$ are $\ket{\downarrow \downarrow}, \ket{\downarrow \uparrow}, \ket{\uparrow \downarrow},$ and $\ket{\uparrow \uparrow}$.

\subsection{Implementing Dephasing Noise} \label{sec:dephasing_noise}

The final element required for our analysis is how we will implement dephasing noise in our qubit estimation procedure. Dephasing, which reduces the coherence of a quantum system, decreases the amplitude of the off-diagonal elements of the state's density matrix, transforming a single qubit state $\rho_0$ into $\rho_1$ as shown,
\begin{align}
	\rho_0 = \begin{pmatrix} \rho_{11} & \rho_{12} \\ \rho_{21} & \rho_{22} \end{pmatrix} \rightarrow \rho_1 = \begin{pmatrix} \rho_{11} & \sqrt{\eta} \rho_{12} \\ \sqrt{\eta} \rho_{21} & \rho_{22} \end{pmatrix}.
\end{align}
The parameter $\eta$ describes how strong the dephasing noise is: $\eta=1$ corresponds to no dephasing, and $\eta=0$ corresponds to total dephasing. Dephasing transforms a single qubit via the formula \cite{DemkowiczDobrzanski2014}
\begin{align}
	\rho_1 = K_0^\eta \rho_0 \left(K_0^\eta\right)^\dagger + K_1^\eta \rho_0 \left(K_1^\eta\right)^\dagger = \begin{pmatrix} \rho_{11} & \sqrt{\eta}\, \rho_{12} \\  \sqrt{\eta}\, \rho_{21} & \rho_{22} \end{pmatrix},
\end{align}
where the operators $K_0^\eta$ and $K_1^\eta$ are two different transformations, defined as
\begin{align}
	K_0^\eta = \left( \frac{1 + \sqrt{\eta}}{2}\right)^{1/2} \begin{pmatrix} 1 & 0 \\ 0 & 1 \end{pmatrix} \\
	K_1^\eta = \left( \frac{1 - \sqrt{\eta}}{2}\right)^{1/2} \begin{pmatrix} 1 & 0 \\ 0 & -1 \end{pmatrix},
\end{align}
and the dagger represents the Hermitian conjugate.

Implementing dephasing on the two qubit state $\rho_{00}$ is done by taking the tensor product of two dephased single qubits, resulting in the formula 
\begin{align}
	\rho_{11} = \sum_{j,l=0,1}K_j^\eta \otimes K_l^\eta \rho_{00} \left(K_j^\eta\right)^\dagger \otimes \left(K_l^\eta\right)^\dagger.
\end{align}
Due to computational constraints, we need to implement the noise discretely in our simulations. To add noise in $n$ steps, we'll first apply the dephasing operator $K^\eta \otimes K^\eta$, rotate by the angle $\phi/n$, and then repeat $n$ times. Note that $\eta$ is not discretized, so the amount of noise added depends both on the dephasing parameter $\eta$ and the number of discrete steps $n$. We can define an operator $\Lambda_{j,l}$, where
\begin{align}†
	\Lambda^\eta_{j, l}(\phi/n) = U_{\phi/n}  \sum_{j,l=0,1}K_j^\eta \otimes K_l^\eta
\end{align}
A single rotation would have the form
\begin{align}
	\rho_{11} = \sum_{j,l=0,1} \Lambda^\eta_{j,l} \rho_{00} \left( \Lambda^{\eta}_{j,l} \right)^\dagger,
\end{align}
and $n$ consecutive rotations would look like
\begin{align}
	\rho_{f} = \sum_{\substack{j_1\cdots j_n=0,1\\ l_1\cdots l_n=0,1}} \left[\prod_{m=1}^n \Lambda^\eta_{j_m, l_m} \rho_i \left(  \prod_{m=1}^n \Lambda^\eta_{j_m, l_m} \right)^\dagger\right]. \label{eq:noisy_rotation}
\end{align}

This gives us all the tools necessary to simulate both noiseless and noisy two-qubit probes being rotated by an angle $\phi$, and then to measure the probes and estimate the parameter $\phi$.

\section{Computational Considerations} \label{sec:comp_considerations}

We will be using simulations to study both noiseless and noisy estimation procedures, so it is in order to note a few computational considerations. First, quantum measurement is inherently random. In order to simulate that randomness, we will use a random number generator to determine the outcome of each measurement, with outcomes weighted by their probabilities. The consequence of this method is that each simulation will be slightly different. To account for this fact, we will run multiple trials of each simulation and study the average behavior of the trials, as described in Sec. \ref{sec:bayesian_quantifier}. 

Second, we will need to sample a large number of values of $\phi$ to determine the confidence interval $L(CI)$. We can use Monte-Carlo methods to compute this quantity. We will begin by randomly sampling the entire domain of $\phi$, then finding the confidence interval nearest to our target confidence level of 95\%. If this confidence interval is outside of a certain tolerance $\tau$ of the 95\% value, then we sample additional angles $\phi$ around the endpoints of the confidence interval in order to get within our tolerance. We then find the uncertainty $L(CI)$ by taking the distance between the two endpoints of the confidence interval. In our simulations we will set the confidence threshold to $\tau = 10^{-3}$.

Third, we repeated each estimation $N_E$ times, as described in Sec. \ref{sec:bayesian_quantifier}. For the noiseless simulations, we used $N_E=10^3$, and for the noisy simulations we used  $N_E=5 \cdot 10^2$.

Finally, we discretely compute the uncertainty averaged over all angles $\phi$, mean($\mu(L(CI))$). To ensure consistency in our averages, we averaged over $N_\phi$ angles evenly spaced between 0 and $\pi/2$. For the noiseless simulations, we used $N_\phi=20$, and for the noisy simulations, we used $N_\phi=10$.

Due to computational constraints, the values of $N_E$ and $N_\phi$ are lower for the noisy simulations, reducing the number of points we average over in the noisy datasets. We tested various values of $N_E$ and $N_\phi$ in the noiseless simulation, and found that decreasing either value make the resulting plots less smooth, but as long as both values were above a certain threshold, $N_E \gtrapprox 100$ and $N_\phi \gtrapprox 5$, the plots did not change qualitatively. As such, the reduced values of $N_\phi$ and $N_E$ in the noisy simulations limit the precision to which we can make statements about the effect of noise on our probes, but we do not believe that they qualitatively alter our results.

\section{Results} \label{sec:results}

In this section we present the results of our simulations. We  use our simulations to compute the uncertainty metric mean($\mu(L(CI))$) for the qubit estimation procedure, as described in Secs. \ref{sec:bayesian_quantifier} and \ref{sec:comp_considerations}. We  perform our simulations using three different levels of dephasing noise: noiseless ($\eta = 1.0$), low noise ($\eta=0.9$), and high noise ($\eta=0.5$). These regimes are qualitatively labeled. We find that in the low noise regime, the quantum advantage is reduced, and in the high noise regime, that advantage is lost all together. 


\subsection{Noiseless Qubit Quantum Metrology Results}

\begin{figure}
	\centering
	\includegraphics[width=\linewidth]{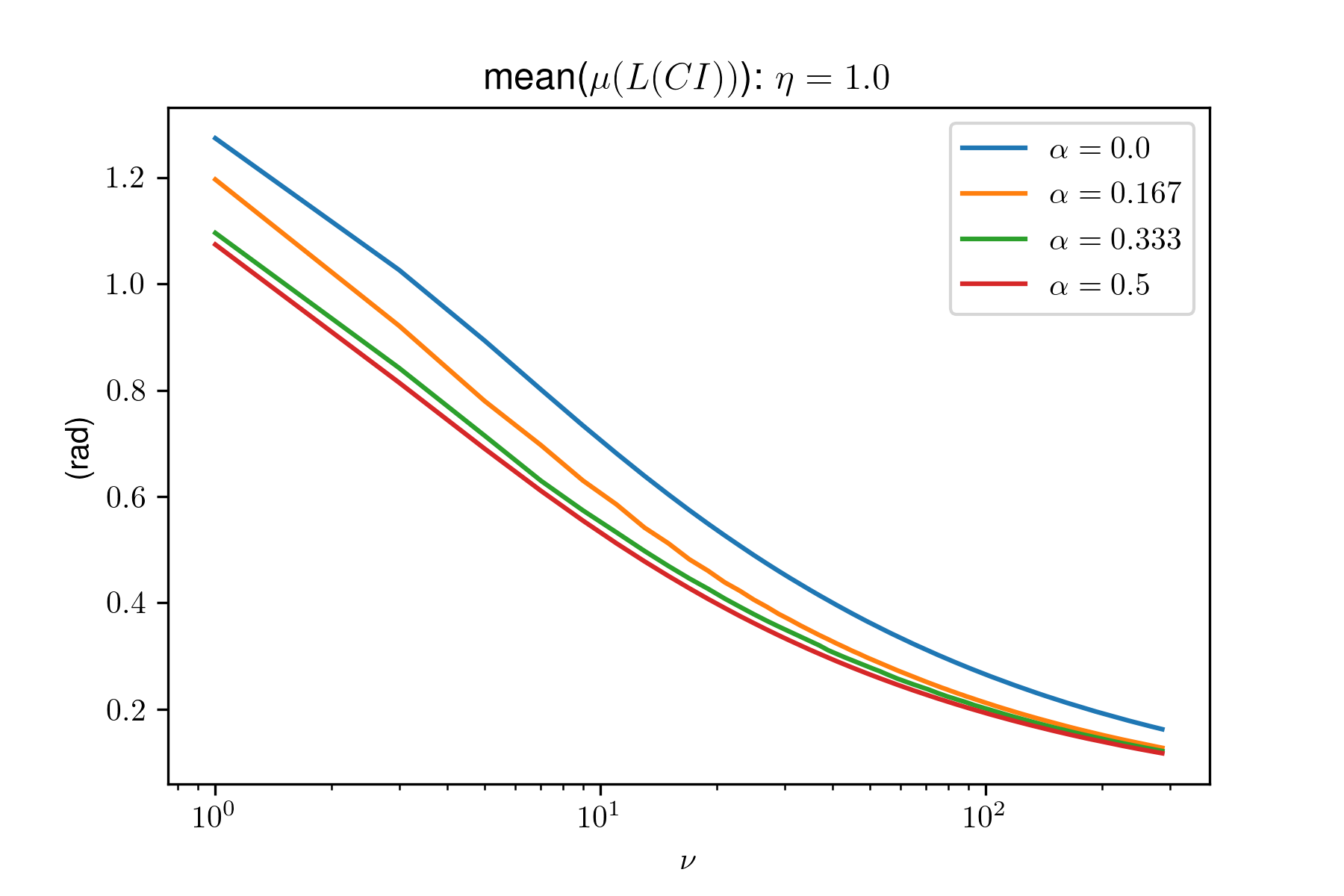}
	\caption[Absolute uncertainty as a function of the number of measurements $\nu$ for the noiseless simulation.]{The uncertainty from the noiseless simulation ($\eta=1.0$) as a function of the number of measurements $\nu$, shown for four different initial states parameterized by $\alpha$, $\ket{\psi} = \sqrt{\alpha} \ket{\downarrow \uparrow} + \sqrt{1-\alpha} \ket{\uparrow \downarrow}$.}
	\label{fig:mean_mu_l_ci}
\end{figure}

\begin{figure}
	\centering
	\includegraphics[width=\linewidth]{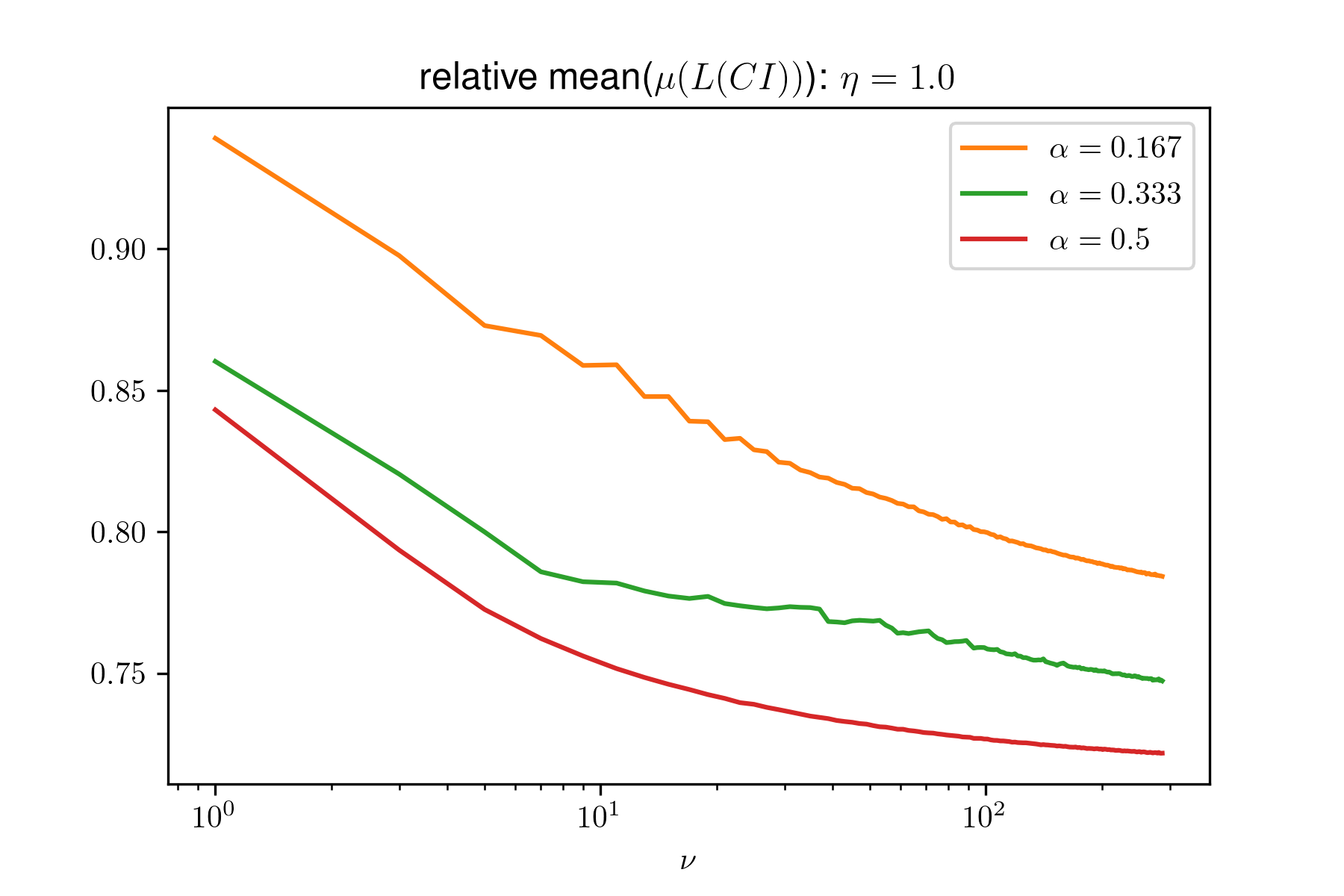}
	\caption[Relative uncertainty as a function of the number of measurements $\nu$ for the noiseless simulation.]{The relative uncertainty from the noiseless simulation ($\eta=1.0$) as a function of the number of measurements $\nu$, shown for three different initial states parameterized by $\alpha$, $\ket{\psi} = \sqrt{\alpha} \ket{\downarrow \uparrow} + \sqrt{1-\alpha} \ket{\uparrow \downarrow}$. The relative uncertainty is the ratio between the uncertainty of a given initial state $\ket{\psi}$ and the uncertainty of a separable state $\ket{\psi} = \ket{\downarrow \uparrow}$, both for the same number of measurements $\nu$.}
	\label{fig:mean_mu_l_ci_rel}
\end{figure}

Our results for the noiseless qubit estimation scheme simulation are given in Figs. \ref{fig:mean_mu_l_ci} and \ref{fig:mean_mu_l_ci_rel}. We  tested the initial probe state $\ket{\psi} = \sqrt{\alpha} \ket{\downarrow \uparrow} + \sqrt{1-\alpha} \ket{\uparrow \downarrow}$ for four different values of $\alpha$: 0, 1/6, 1/3, and 1/2. When $\alpha=0$, the probe is separable, and when $\alpha=1/2$, the probe is a maximally entangled Bell state. The closer $\alpha$ is to $1/2$, the more entangled the probe is \cite{Horodecki2009Negativity}. Figure \ref{fig:mean_mu_l_ci} gives the uncertainty for each state as a function of $\nu$, the number of measurements used in the estimation. Figure \ref{fig:mean_mu_l_ci_rel} gives the relative uncertainty for the three entangled probe states, defined as the uncertainty of the entangled probe divided by the uncertainty of the separable probe for the same number of measurements $\nu$. This relative uncertainty allows us to see if the entangled probe state is performing better (relative uncertainty less than one) or worse (relative uncertainty greater than one) than the separable probe state.

Equation (\ref{eq:relative_qcrb}) from Sec. \ref{sec:introduction} tells us that as the number of measurements $\nu$ becomes asymptotically large, the relative uncertainty for the maximally entangled state should approach $1/\sqrt{2} \approx 0.71$. We  are using our own uncertainty metric, but this claim still seems to hold: Fig. \ref{fig:mean_mu_l_ci_rel} shows that the relative uncertainty for the state $\alpha=0.5$ seems to be approaching an asymptote in the region of 0.71, although we have insufficient data to state this conclusively. However, as the number of measurements $\nu$ decreases, the relative uncertainty increases. Apparently, the advantage of using the entangled states decreases as the number of measurements $\nu$ decreases. A single maximally entangled probe gives an uncertainty about $15\%$ smaller than a single separable probe, but ten maximally entangled probes give an uncertainty about $25\%$ smaller than ten separable probes.


\subsection{Noisy Qubit Quantum Metrology Results}

\begin{figure}
	\centering
	\includegraphics[width=\linewidth]{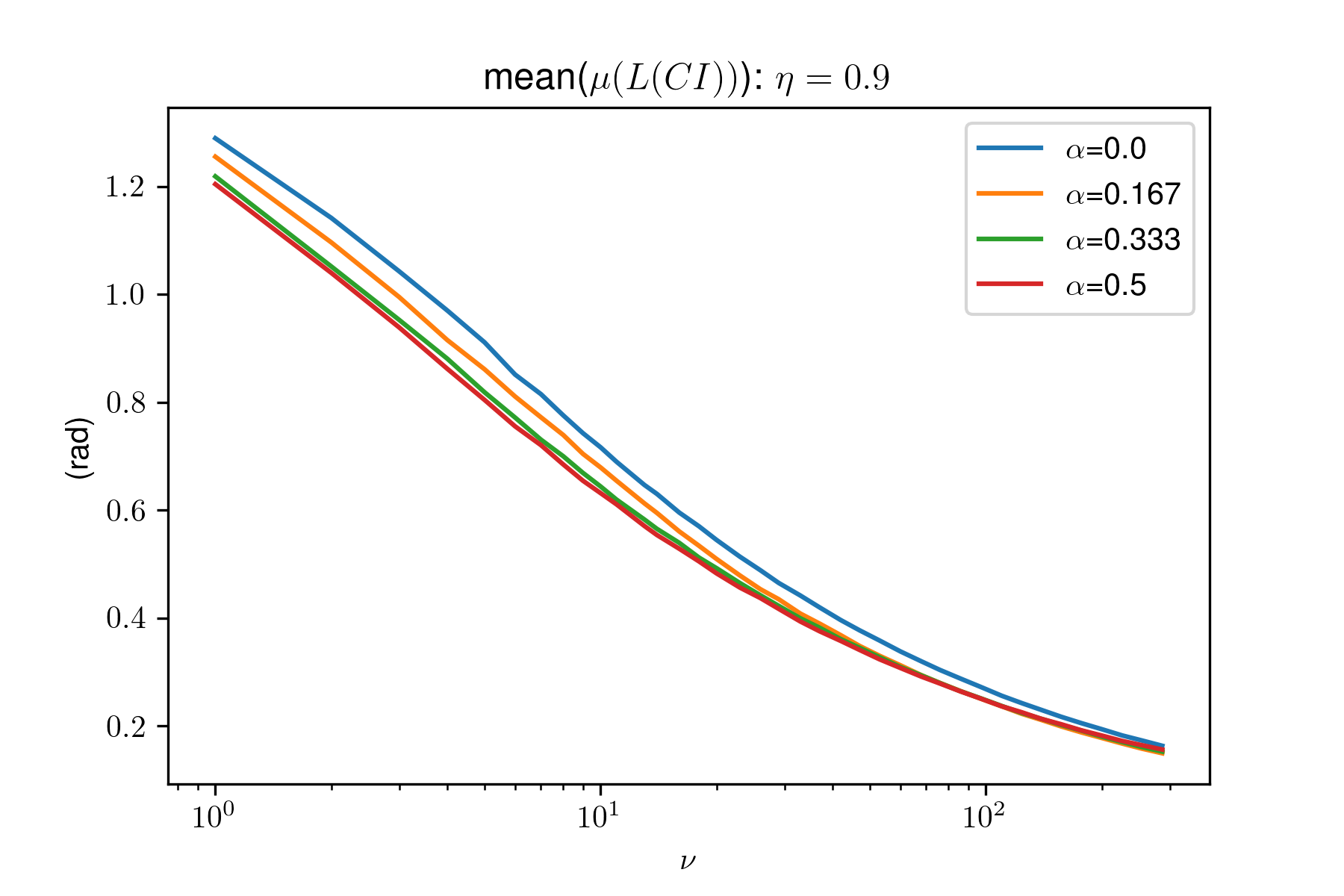}
	\caption[Absolute uncertainty as a function of the number of measurements $\nu$ for the low noise simulation.]{The uncertainty from the low noise simulation ($\eta=0.9$) as a function of the number of measurements $\nu$, shown for four different initial states parameterized by $\alpha$, $\ket{\psi} = \sqrt{\alpha} \ket{\downarrow \uparrow} + \sqrt{1-\alpha} \ket{\uparrow \downarrow}$.}
	\label{fig:mean_mu_l_ci_09}
\end{figure}

\begin{figure}
	\centering
	\includegraphics[width=\linewidth]{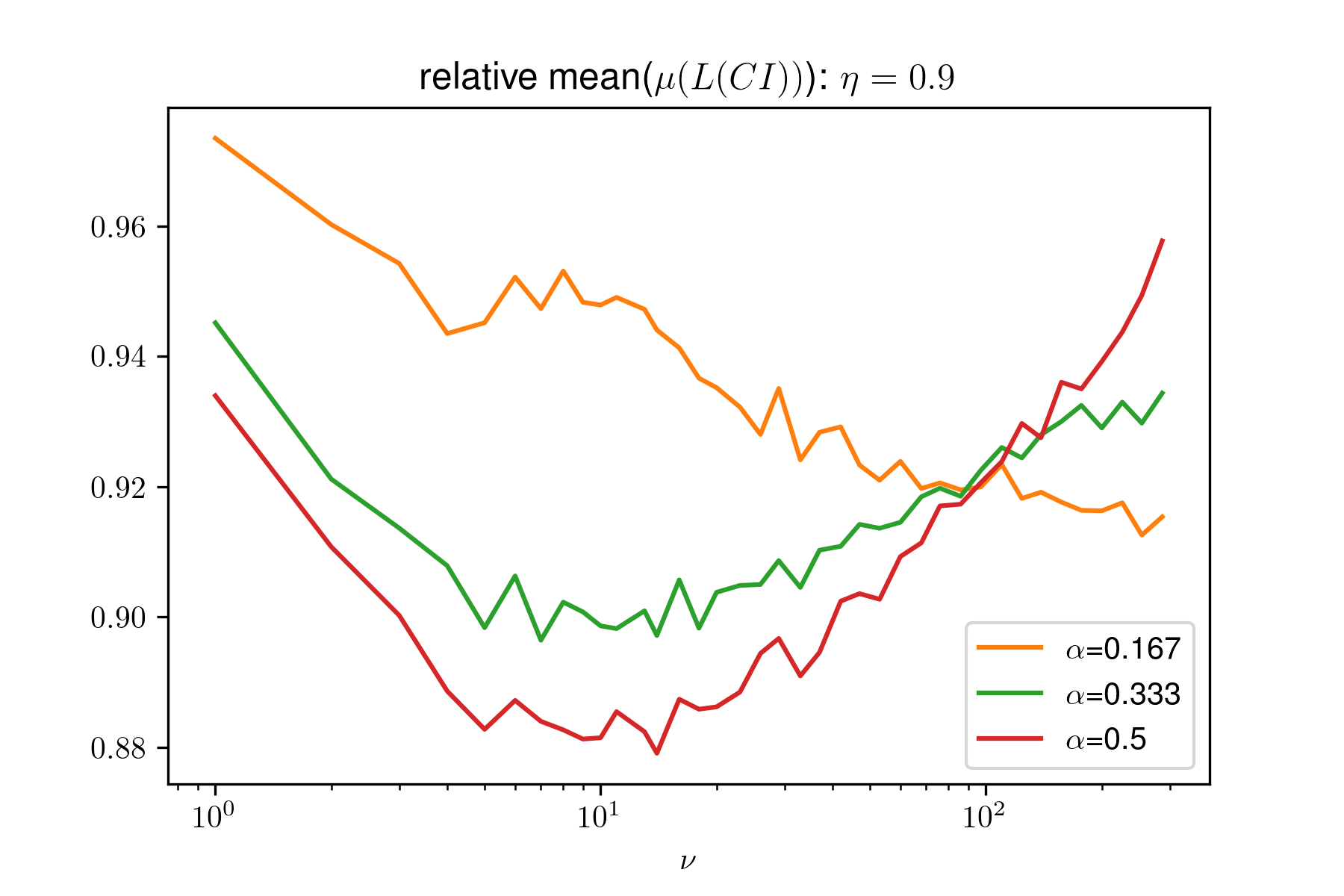}
	\caption[Relative uncertainty as a function of the number of measurements $\nu$ for the low noise simulation.]{The relative uncertainty from the low noise simulation ($\eta=0.9$) as a function of the number of measurements $\nu$, shown for three different initial states parameterized by $\alpha$, $\ket{\psi} = \sqrt{\alpha} \ket{\downarrow \uparrow} + \sqrt{1-\alpha} \ket{\uparrow \downarrow}$. The relative uncertainty is the ratio between the uncertainty of a given initial state $\ket{\psi}$ and the uncertainty of a separable state $\ket{\psi} = \ket{\downarrow \uparrow}$, both for the same number of measurements $\nu$.}
	\label{fig:mean_mu_l_ci_09_rel}
\end{figure}

\begin{figure}
	\centering
	\includegraphics[width=\linewidth]{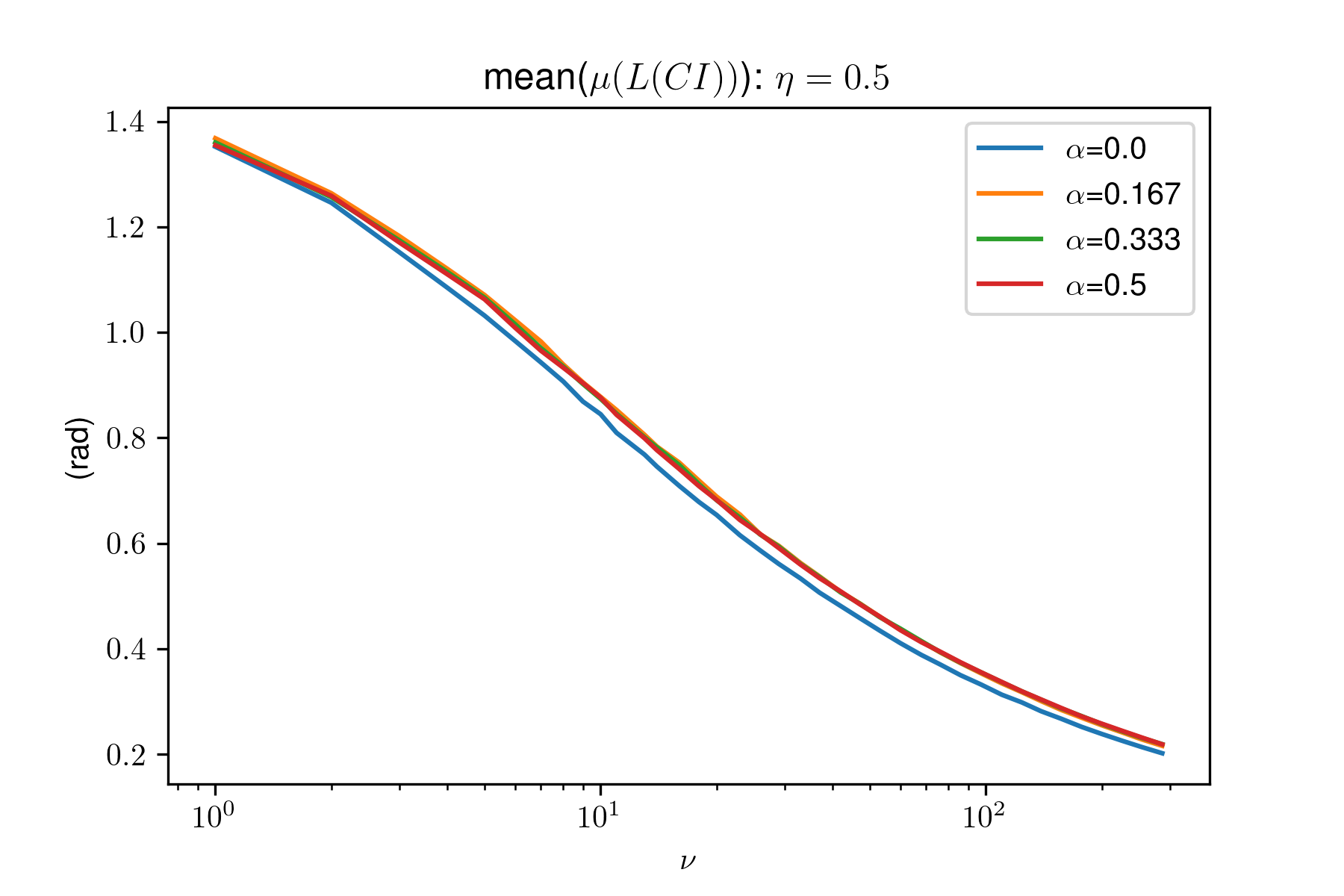}
	\caption[Absolute uncertainty as a function of the number of measurements $\nu$ for the high noise simulation.]{The  uncertainty from the high noise simulation ($\eta=0.5$) as a function of the number of measurements $\nu$, shown for four different initial states parameterized by $\alpha$, $\ket{\psi} = \sqrt{\alpha} \ket{\downarrow \uparrow} + \sqrt{1-\alpha} \ket{\uparrow \downarrow}$.}
	\label{fig:mean_mu_l_ci_05}
\end{figure}

\begin{figure}
	\centering
	\includegraphics[width=\linewidth]{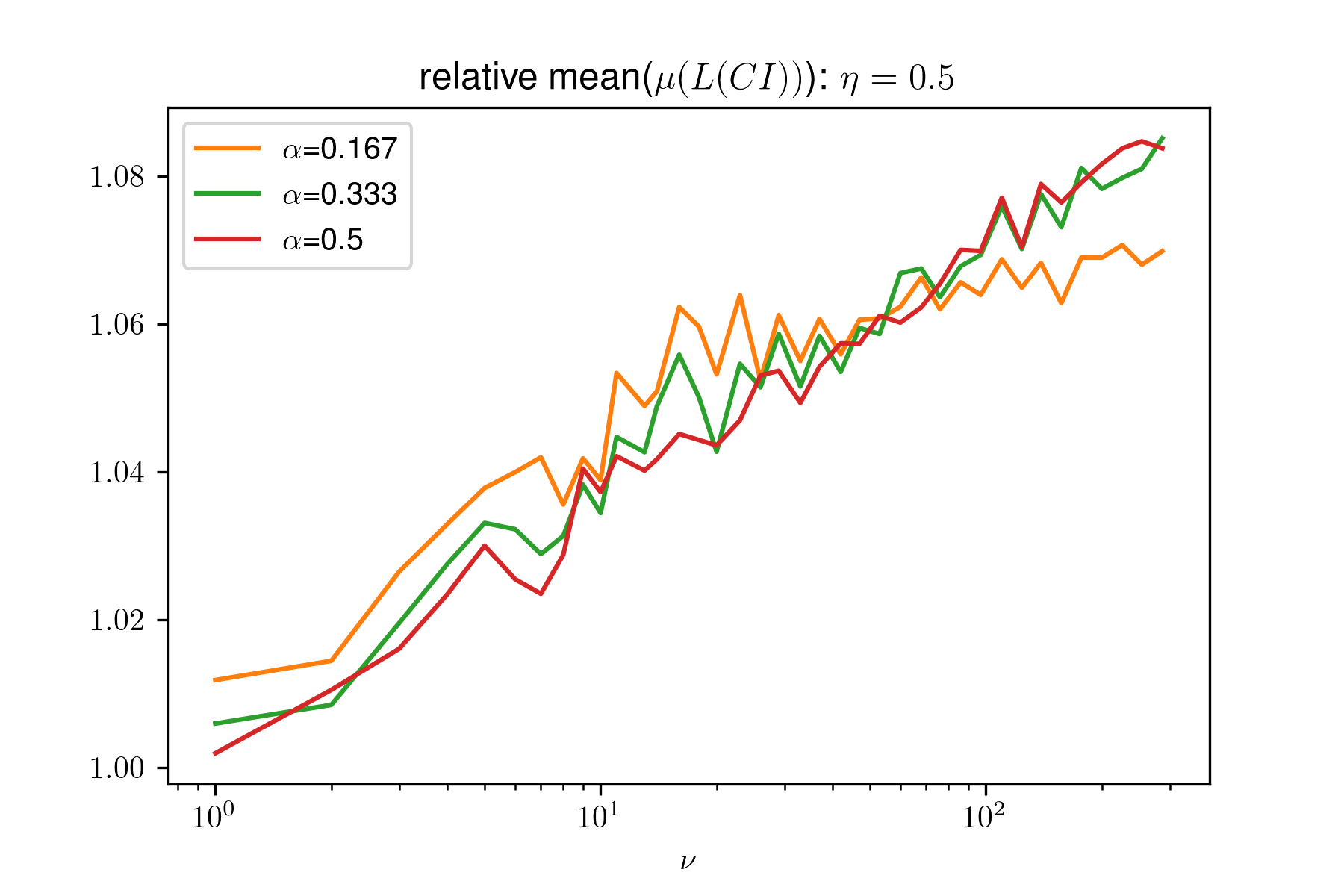}
	\caption[Relative uncertainty as a function of the number of measurements $\nu$ for the high noise simulation.]{The relative uncertainty from the high noise simulation ($\eta=0.5$) as a function of the number of measurements $\nu$, shown for three different initial states parameterized by $\alpha$, $\ket{\psi} = \sqrt{\alpha} \ket{\downarrow \uparrow} + \sqrt{1-\alpha} \ket{\uparrow \downarrow}$. The relative uncertainty is the ratio between the uncertainty of a given initial state $\ket{\psi}$ and the uncertainty of a separable state $\ket{\psi} = \ket{\downarrow \uparrow}$, both for the same number of measurements $\nu$.}
	\label{fig:mean_mu_l_ci_05_rel}
\end{figure}

In this section, we consider the effect of dephasing noise on the qubit estimation procedure. We use the procedure described in Sec. \ref{sec:dephasing_noise}. The dephasing is implemented in $n=5$ discrete steps, and the total amount of noise is parameterized by both $\eta$ and $n$. For our case of $n=5$, the parameter $\eta=0.9$ corresponds to low levels of dephasing, and $\eta=0.5$ corresponds is high levels of dephasing.

Our results are given for the two cases in Figs. \ref{fig:mean_mu_l_ci_09}-\ref{fig:mean_mu_l_ci_05_rel}. Figures \ref{fig:mean_mu_l_ci_09} and \ref{fig:mean_mu_l_ci_05} show the results in absolute form, and Figs. \ref{fig:mean_mu_l_ci_09_rel} and \ref{fig:mean_mu_l_ci_05_rel} show the results in relative form, where the ratio between the uncertainty of the entangled states and the separable state is plotted.

We consider first the low noise case. Figure \ref{fig:mean_mu_l_ci_09} shows that, for all four values of $\alpha$, increasing the number of measurements $\nu$ decreases the uncertainty of the estimation. Additionally, Fig. \ref{fig:mean_mu_l_ci_09_rel} shows that for all values of $\nu$ plotted, the entangled states outperform the separable state. For low $\nu$, the more entangled a state is, the better it outperforms the separable state. 

Notice, however, that in comparison to the noiseless simulation in Fig. \ref{fig:mean_mu_l_ci_rel}, the quantum advantage is smaller in the low-noise simulation--the relative uncertainty for the maximally entangled state at $\nu=10$ is equal to approximately 0.88 for the low-noise simulation, whereas it was equal to about 0.75 for the noiseless simulation. 

Furthermore, as $\nu$ increases, the less entangled states gradually outperform the maximally entangled state, and the relative uncertainty for $\alpha=0.5$ and $\alpha=0.333$ seem to be trending towards one. We  expect that if a wider range of $\nu$ values were plotted, we would see the same behavior for the $\alpha=0.167$ state as well.

In the case of high noise, Fig. \ref{fig:mean_mu_l_ci_05} shows that the uncertainty of the estimation decreases as the number of measurements $\nu$ increases for all four values of $\alpha$. Notice, though, that the relative plot in Fig. \ref{fig:mean_mu_l_ci_05_rel} shows that the relative uncertainty is greater than one for all values of $\nu$. Not only is there no quantum advantage, but there even seems to be a quantum disadvantage. Within the accuracy of our simulation, there doesn't seem to be a definitive difference between the three entangled states, although the state $\alpha=0.167$ does seem to outperform the other two at large enough $\nu$ values.

In summary, both of our noisy simulations showed that the uncertainty of the estimation decreases as the number of measurements increases, regardless of the probe state used. With regards to the quantum advantage, our low noise simulations indicate that it does still exist, although it is smaller than in the noiseless case. As $\nu$ increases, less entangled states begin to outperform the maximally entangled state, and the quantum advantage appears to vanish all together as $\nu$ becomes sufficiently large. 

In the case of high noise, there is no quantum advantage for any value of $\nu$, and in fact there appears to be a quantum disadvantage. Our simulations indicate no clear distinction between the three entangled states in the presence of high noise. 

It is worth reiterating that our model added dephasing in five discrete steps. In a physical experiment, we would expect noise to affect the probes continuously, so further work would be in order to determine which of these features hold as noise is added to the probes in smaller increments.

\section{Conclusion} \label{sec:conclusion}

Quantum metrology shows how quantum resources, such as entanglement, can decrease the uncertainty of a parameter estimation. If no noise is present and asymptotically many measurement resources are available, then this phenomenon is well described by the Quantum Cramer-Rao Bound \cite{BraunsteinCaves1994}. However, there is no general description of the quantum advantage in the regime of limited measurement resources. 

We have defined a Bayesian uncertainty quantifier appropriate to this regime of limited resources, and developed the mathematical description necessary to simulate a qubit metrology scheme. We  then used simulations to study qubit probes estimating a rotation angle $\phi$. We  simulated both noiseless and noisy estimation schemes. Our noiseless results show that, for the class of states we tested, entanglement always decreases the uncertainty of an estimation procedure, although the quantum advantage is reduced when fewer probes are used. Our noisy results, on the other hand, show that for small amounts of dephasing noise, the entangled states outperformed the separable state for the range of $\nu$ plotted, although it appears that after a given number of measurements $\nu$ the separable state will outperform the entangled states. In the presence of large amounts of dephasing noise, the separable state outperforms the entangled states for all values of $\nu$.

It is important to note that our results are not general. We tested a specific class of states, $\ket{\psi} = \sqrt{\alpha} \ket{\downarrow \uparrow} + \sqrt{1-\alpha} \ket{\uparrow \downarrow}$, and our conclusions apply only to that class of states. We have performed preliminary work that indicates that our class of states is representative of a larger space of quantum states, but further work needs to be done to extend these conclusions to all qubit states.

\section{Acknowledgements}

We acknowledge support from the College of Physical and Mathematical Sciences at Brigham Young University.

\bibliography{../thesis_js}

\end{document}